\pdfoutput=1

\documentclass[11pt]{article}

\usepackage{acl}

\usepackage{times}
\usepackage{latexsym}

\usepackage{amsmath}
\usepackage{algorithm}
\usepackage{algorithmic}
\usepackage{enumitem}
\usepackage{xspace}
\usepackage{booktabs}
\usepackage{graphicx}

\newcommand{\tool}[0]{\textsf{{MEGAnno}}\xspace}
\newcommand{\username}{Meggie\xspace} 

\graphicspath{{fig/}{./}}

\usepackage{listings}
\usepackage{xcolor}
\definecolor{codegreen}{rgb}{0,0.6,0}
\definecolor{codegray}{rgb}{0.5,0.5,0.5}
\definecolor{codepurple}{rgb}{0.58,0,0.82}
\definecolor{backcolour}{rgb}{0.95,0.95,0.92}

\lstdefinestyle{mystyle}{
    backgroundcolor=\color{backcolour},   
    commentstyle=\color{codegreen},
    keywordstyle=\color{magenta},
    numberstyle=\tiny\color{codegray},
    stringstyle=\color{codepurple},
    basicstyle=\ttfamily\footnotesize,
    breakatwhitespace=false,         
    breaklines=true,                 
    captionpos=b,                    
    keepspaces=true,                 
    numbers=left,                    
    numbersep=5pt,                  
    showspaces=false,                
    showstringspaces=false,
    showtabs=false,                  
    tabsize=2
}

\usepackage[T1]{fontenc}

\usepackage[utf8]{inputenc}

\usepackage{microtype}

\title{\tool: Exploratory Labeling for NLP in Computational Notebooks}

\author{Dan Zhang\thanks{~~~Equal contribution.}~, Hannah Kim\footnotemark[1]~, Rafael Li Chen, Eser Kandogan, Estevam Hruschka\\
    Megagon Labs \\
    \texttt{\{dan\_z,hannah,rafael,eser,estevam\}@megagon.ai}
}

\begin{document}
\maketitle

\begin{abstract}
We present \tool, a novel exploratory annotation framework designed for NLP researchers and practitioners.
Unlike existing labeling tools that focus on data labeling only, our framework aims to support a broader, iterative ML workflow including data exploration and model development.
With \tool's API, users can programmatically explore the data through sophisticated search and automated suggestion functions and incrementally update labeling schema as their project evolve. 
Combined with our widget, the users can interactively sort, filter, and assign labels to multiple items simultaneously in the same notebook where the rest of the NLP project resides. 
We demonstrate \tool's flexible, exploratory, efficient, and seamless labeling experience through a sentiment analysis use case.
\end{abstract}

\section{Introduction}
\label{sec:intro}
Data labeling is an important step in the machine learning life cycle since the quality and quantity of training data directly affect the model performance~\cite{garbage2021}.
Unfortunately, existing annotation tools tend to consider the data labeling step in isolation from the broader ML life cycle, ignoring the iterative workflow of researchers and practitioners.
However, activities such as data selection, exploratory data analysis, data labeling, model training, and evaluation do not happen sequentially~\cite{rahman2022ie}. 
Instead, continuous iterations are required to improve data, annotation, and models~\cite{hohman2020iterativeML}.

\begin{figure}[t]
\includegraphics[width=\columnwidth]{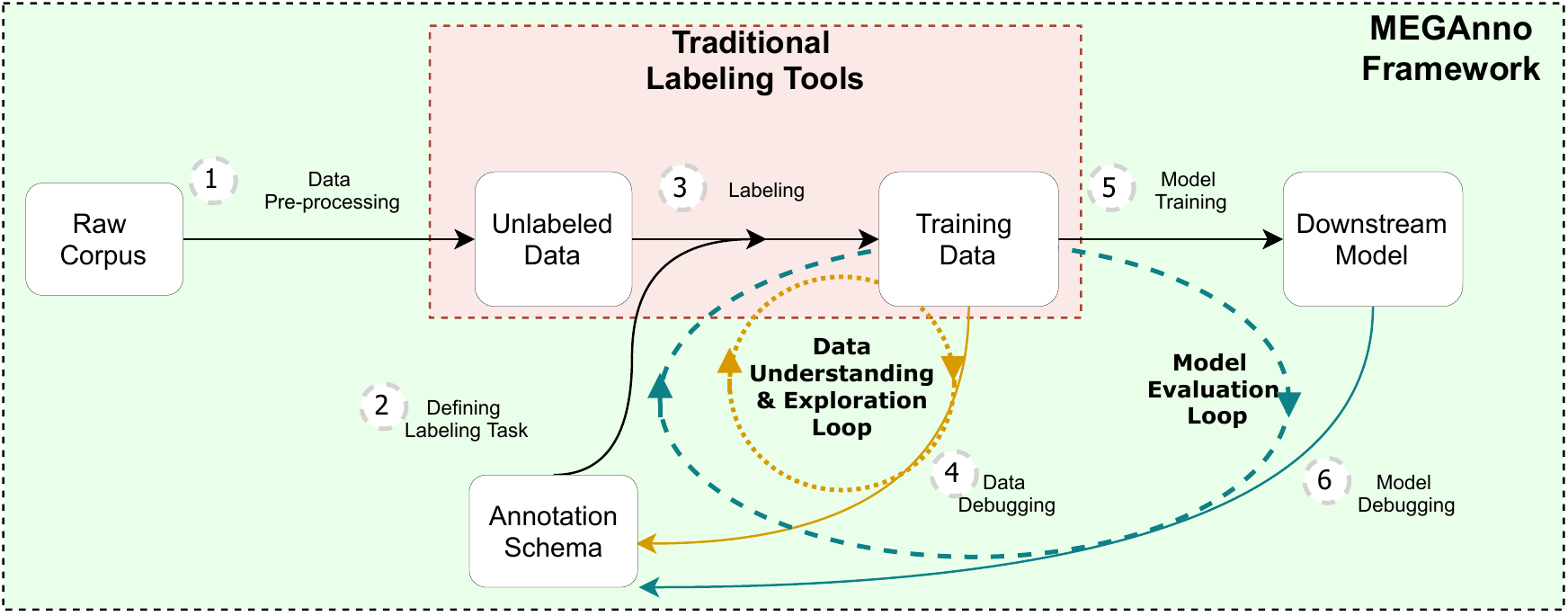}
\caption{
Dual-loop model for data annotation: 
(1) data understanding/exploration loop (yellow): iteratively update annotation schema while exploring and annotating data 
and (2) model evaluation loop (green): train and improve a downstream model over iterations by debugging data.
Most tools are focused on the labeling step only (red box). 
\tool aims to capture both loops seamlessly within the framework (green box).}
\label{fig:dual-loop}
\end{figure}

To further investigate this gap, we examine the data annotation practices
within the ML model development life cycle.
Based on a formative study with six researchers from our organization, we characterize their annotation practices as a ``dual-loop'' model shown in Fig.~\ref{fig:dual-loop}. 
After data preprocessing (Fig.~\ref{fig:dual-loop} {\large \textcircled{\small 1}}), researchers define their annotation schema in terms of what labels to collect, how many data points are needed, and so on (Fig.~\ref{fig:dual-loop} {\large \textcircled{\small 2}}).
As they explore and annotate the data (Fig.~\ref{fig:dual-loop} {\large \textcircled{\small 3}}), they often go back and refine the annotation schema due to their improved understanding of the data and updated mental models for the tasks (Fig.~\ref{fig:dual-loop} {\large \textcircled{\small 4}}). 
For example, in a document classification task, a user may start with loosely defined category labels and add more choices as she discovers relevant documents~\cite{Felix18}. 
Throughout this paper, we refer to this cycle as \textbf{data understanding/exploration loop}.
Next, the labeled data is exported from the annotation tool and is used to train a model (Fig.~\ref{fig:dual-loop} {\large \textcircled{\small 5}}).
However, in practice, ML model training is rarely completed in one pass and usually goes through many iterations of labeling, training, data, and model debugging~\cite{pustejovsky2012natural}.
Fig.~\ref{fig:dual-loop} {\large \textcircled{\small 6}} refers to the cases where researchers may need to collect more data (e.g., for the less represented classes due to a sub-optimal prediction performance of the downstream model) or further refine the schema.
We call this cycle (i.e., model training, evaluating and debugging, collecting more data, and training again) \textbf{model evaluation loop}.

We find that the iterative dual-loop workflow of NLP researchers and practitioners is rarely supported in most existing tools.
More specifically, we identified three main challenges towards supporting the full annotation life cycle:

\begin{itemize}
\setlength{\itemsep}{-.3\baselineskip}
\item \textbf{Gaps between ML toolings.} Most of the existing tools are standalone and designed for a specific ML step, which results in frequent context switching and data migration overhead in the researchers' daily workflow.

\item \textbf{Lack of customizable and granular control.} Not all data points are equally important. There are often cases where users might want to prioritize a particular batch (e.g., to achieve better class or domain coverage or focus on the data points that the downstream model cannot predict well). Although some recent active learning based tools~\cite{Prodigy:2018,huamnloop} can provide suggestions for the next batch, most tools do not offer customizable and fine-grained control with or with a downstream model (i.e., covering both loops).

\item \textbf{Lack of support for project evolution.} Current annotation tools usually work with the assumption that the data collection task is well-defined and immutable and ignore that annotation projects can evolve as explorations happen and insights are gathered. Thus they lack the support to help users make evolution decisions, and their immutable nature makes it hard to apply these changes. 
\end{itemize}

To address the challenges, we present \tool, a flexible, exploratory, efficient, and seamless labeling framework for NLP researchers and practitioners. It provides a seamless experience where data pre-processing, annotation, analysis, model development and evaluation can happen in the same notebook, a popular daily working environment for data science practitioners. \tool provides customizable interfaces to help users drive their project to the desired directions through rich heuristic-based search, automatic suggestion, and active learning based suggestions of the next data batch. With project evolution in mind, \tool is designed to work with flexible task schema and provides a built-in analysis dashboard to aid decision-making.
To our knowledge, \tool is the first flexible, exploratory labeling framework that can support ML workflow seamlessly in computational notebooks (Fig.~\ref{fig:dual-loop}: green box). 

\section{\tool}
\label{sec:system}
\subsection{Framework Overview}
\label{sec:overview}

\begin{figure}[h]
\includegraphics[width=\columnwidth]{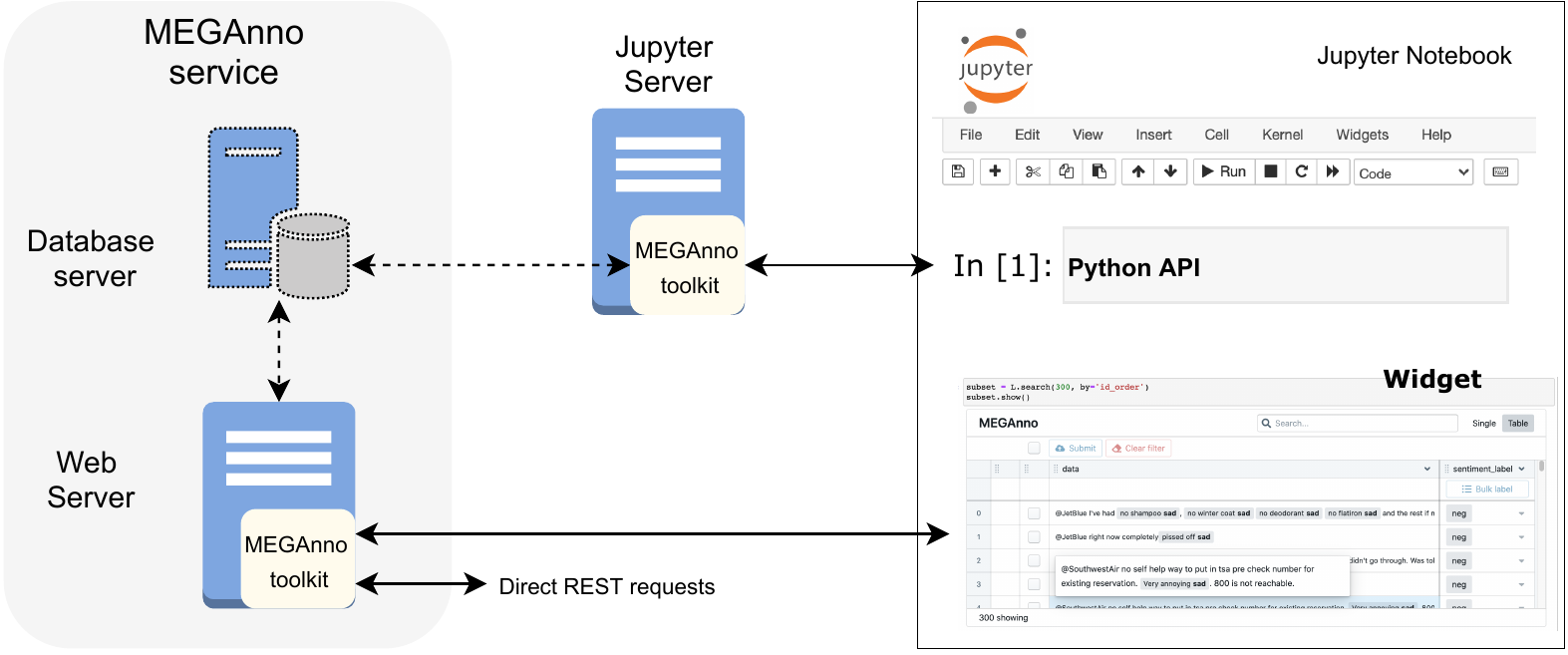}
\caption{The \tool framework provides exploratory annotation services through a toolkit (installable as Python libraries) providing programmatic interfaces, a web server providing language-agnostic REST APIs, and an internal database to store data, annotation, and related artifacts. Solid lines show programmatic interactions through Python APIs calls and REST calls delegated by our notebook widget or directly issued by authenticated applications. Dotted lines show internal interfaces where \tool toolkit handles communication with the database and are hidden from the users.}
\label{fig:diagram-sys}
\end{figure}

\tool provides service through 1) an internal database that stores the data, annotations, and various artifacts produced in the annotation process, 2) a \tool toolkit that provides python API for programmatic and visual data exploration and labeling, and 3) a web server that provides language-agnostic REST APIs (Fig \ref{fig:diagram-sys}). 
After installing the toolkit on a Jupyter server, users will have access to our Python APIs and React-based widget to manage their project, explore and annotate their data from any connected notebook. 

\vspace{-0.7em}
\paragraph{Data model}
A \texttt{Data record} refers to an item in the pre-processed data corpus for labeling. It can be a sentence, a paragraph, a document, or a flattened text from multiple texts such as a question-answer pair. A \texttt{Label} is the smallest unit of user labeling output. \tool currently supports record-level (e.g., topics for document or sentence) and span-level (e.g., named entities) labels. An \texttt{Annotation} is a set of labels given to a data record by an annotator.
\texttt{Metadata} refers to additional information related to the content of a data record (e.g., externally generated part-of-speech tags, embeddings) or of an annotation (e.g., time taken to label, disagreement among annotators). Such information can be helpful in various steps of the ML iterations.
A \texttt{Subset} is a slice of the data records in the database. Subsets can be of random data records or can be generated through search queries that match certain characteristics. 

\paragraph{Task schema}
We support a wide variety of tasks through our customizable schema in JSON format. 
To collect a label, users need to specify the level (i.e., record or span) and provide a list of options to choose from. 
For a task, \tool supports arbitrary numbers of both types of labels. 
We'll see concrete schema examples for a sentiment prediction and extraction task in Section \ref{sec:usecase}. 
At any stage, users can always update the schema to reflect the evolution of the project. There are certain constraints to schema updates to keep the consistency of data. Adding new labels or new label options will always be allowed. Removal of labels and options will trigger a database query and \tool will warn the user if there exist such labeled instances. \footnote{\tool provides an option to clean up legacy labels and retry automatically.}

\subsection{\tool Jupyter Notebook Widget }
\label{sec:frontend}

\tool's interactive notebook widget features 1) our novel table view to facilitate exploratory and batch labeling and 2) the single view, which is similar to traditional labeling UIs.

\paragraph{Table view}
\begin{figure}[t]
\includegraphics[width=\columnwidth]{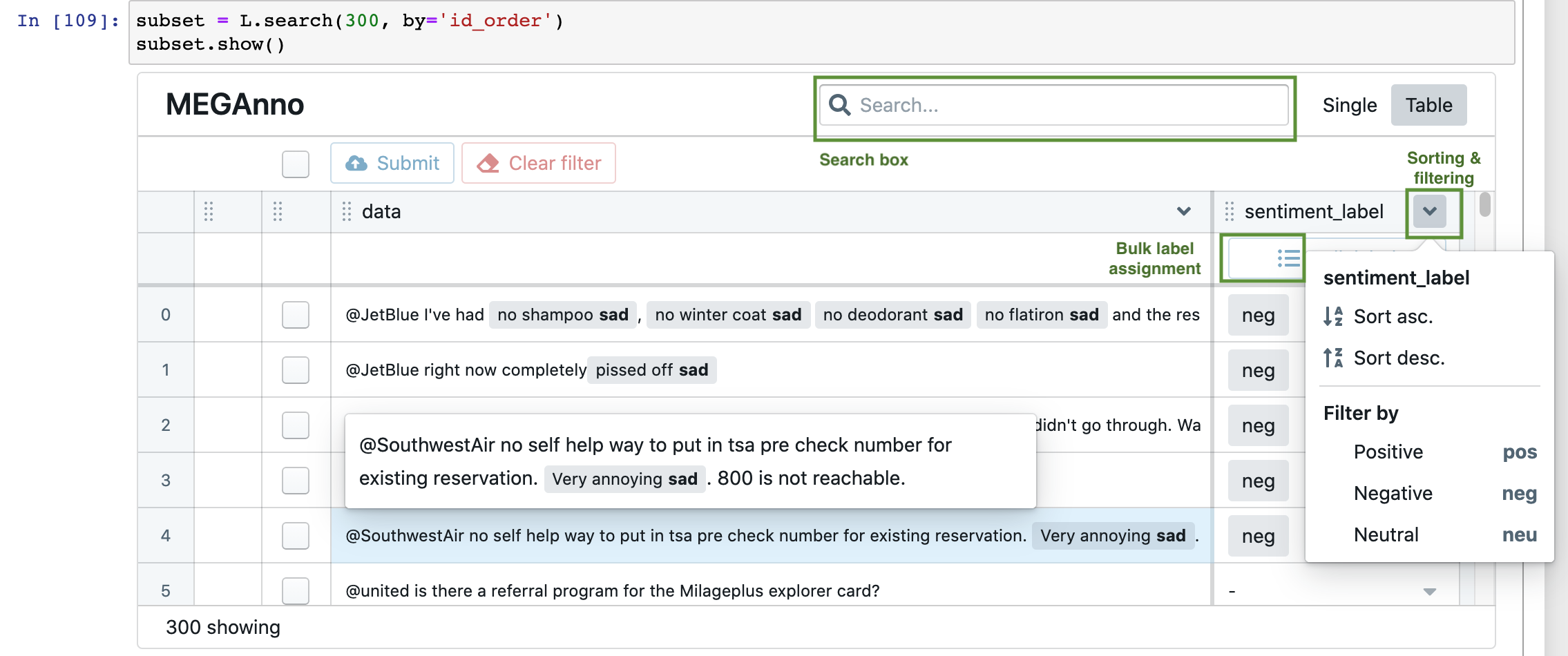}
\caption{The table view to show multiple data records. Hovering over a data record shows its full text in a pop-up. This view allows exploration by searching, sorting, and filtering over labels and single/bulk annotation.}
\label{fig:widget_table_sys}
\end{figure}

The table view (Fig.~\ref{fig:widget_table_sys}) shows data items in a \texttt{Subset} and their annotations if any. 
Each record-level label is shown as a column, and span-level labels are shown together with the highlighted textual span in the data column. 
Users can hover over an item to see full text in a pop-up.
The search box supports three types of search (exact, fuzzy, and regex-based) to filter the data subset further. 
Users can sort and filter rows based on any record-level labels using the dropdown menu.
To assign a record-level label, the users can click on the cell's arrow button and select from the drop-down options. 
Alternatively, the users can assign the same label to multiple records simultaneously by selecting those records and clicking the bulk label button. 

\begin{figure}[h]
\includegraphics[width=\columnwidth]{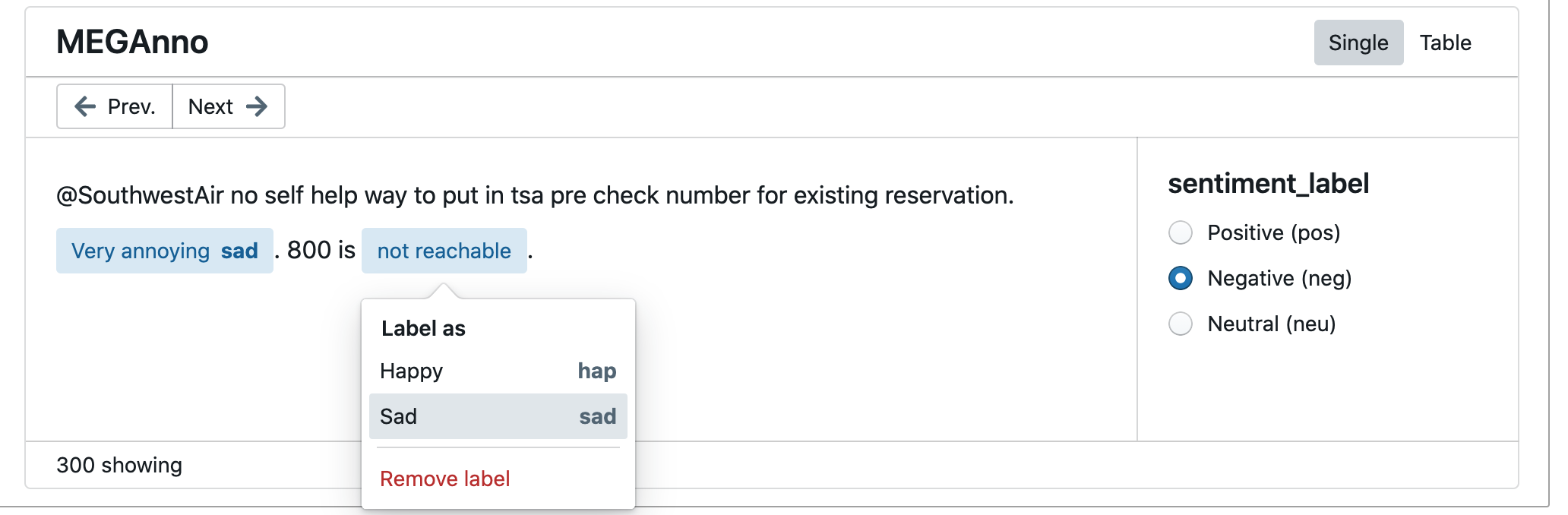}
\caption{The single view to annotate data one by one. In this view, users can drag and label spans for extraction tasks.}
\label{fig:widget_single_sys}
\end{figure}

\paragraph{Single view}
The table view is good for exploration, but the limited space makes span-level annotation cumbersome. So we also provide a single view where users can have a more zoomed-in experience. By clicking the ``Single'' button on the top-right corner or double-clicking on a data record, the widget switches to the single view (Fig.~\ref{fig:widget_single_sys}). In this view, users can assign record-level labels on the right side and span-level labels on the left side by 
selecting/dragging target spans and choosing the label from the options drop-down. Users can loop through the subset using the prev/next button based on the order specified in the table view. At any time, users can switch between the two views by clicking the top right buttons, and the widget preserves all uncommitted annotations during view changes.

\begin{figure}[h]
\includegraphics[width=\columnwidth]{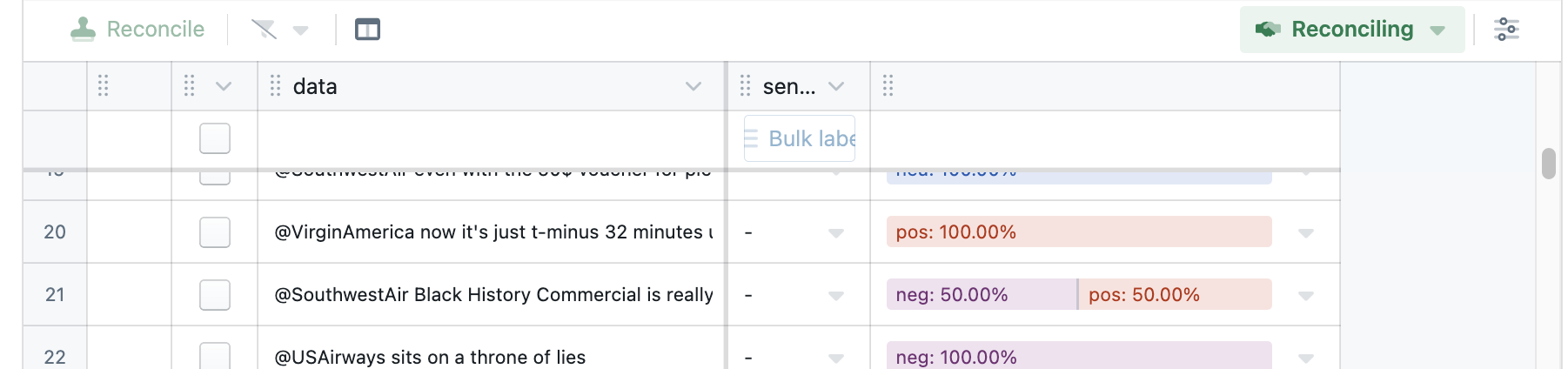}
\caption{Reconcilation view showing the existing label distribution for data points. }
\label{fig:reconcile}
\end{figure}

\paragraph{Working with multiple annotators}
Annotation is rarely done by a single person. As an initial step towards collaborative annotation, \tool provides virtually separated namespaces for each annotator. Users identify themselves by a unique authentication token while connecting to the service and only update their own labels through the widgets. \tool provides a reconciliation view (Fig.~\ref{fig:reconcile}) to look at labels from different individuals and resolve potential conflicts.

\paragraph{Dashboard}

\begin{figure}[h]
\includegraphics[width=\columnwidth]{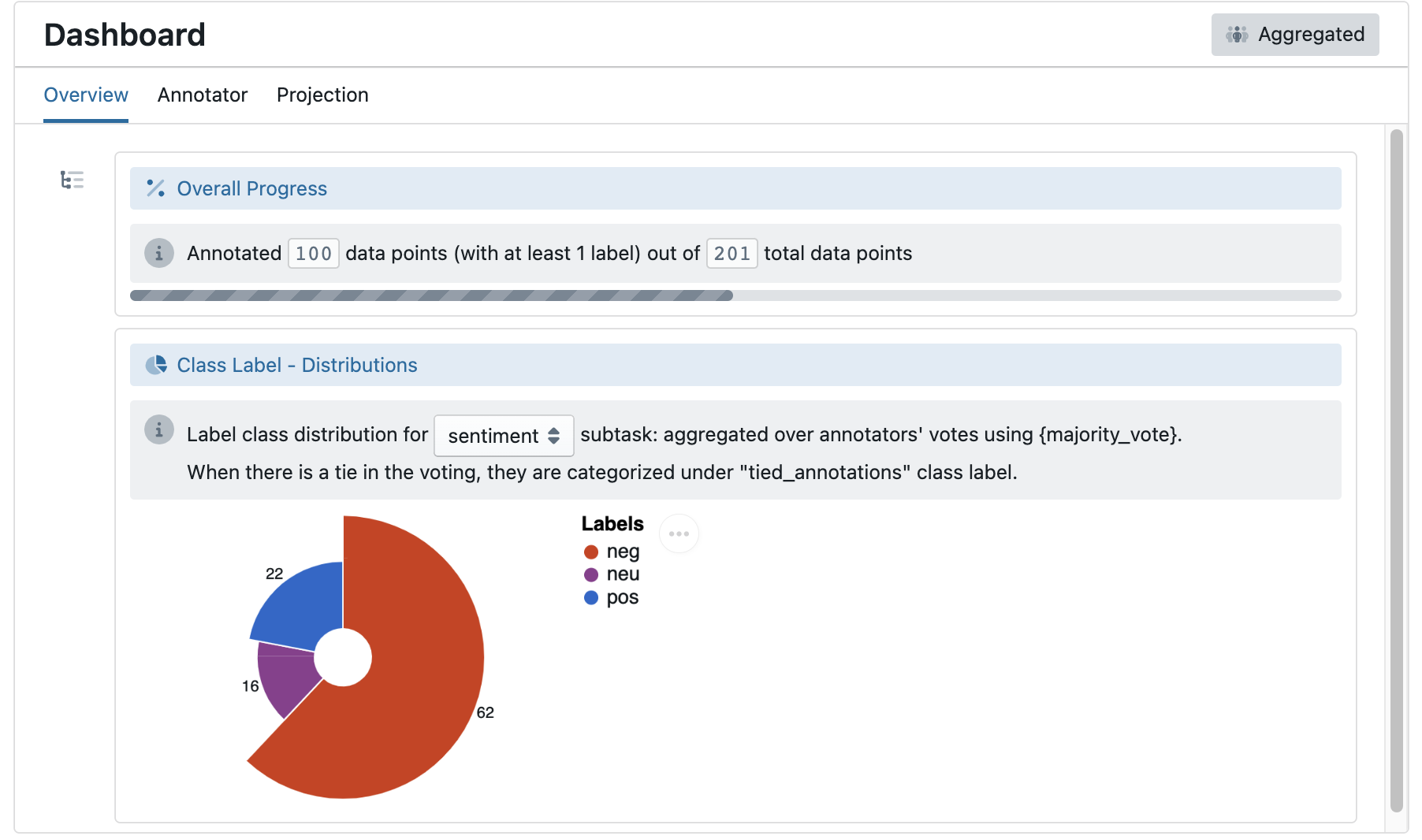}
\caption{Dashboard widget to monitor the progress and statistics of the project and aid decision-making.}
\label{fig:dashboard}
\end{figure}
\tool also provides a built-in visual monitoring dashboard (Fig.~\ref{fig:dashboard}). As projects evolve, users would need to understand the project's status to make decisions about the next steps, like collecting more data points with certain characteristics or adding a new class to the task definition. To aid such analysis, the dashboard widget packs common statistics and analytical visualizations based on a survey of our pilot users. The ``overview'' panel shows statistics about overall progress and per-label class distribution. If multiple annotators are involved, the distribution reflects the majority vote over annotators \footnote{Users can provide their aggregation function to resovle conflicts between annotators}. The remaining ``annotator'' and ``projection'' panels are hidden due to space limitations. To help identify problematic annotators, the annotator pannel shows statistics like overall individual contribution and disagreement scores with others. The projection panel provides customizable visualization to project data points to a two-dimensional visual space. By default, we show the t-SNE~\cite{van2008visualizing} projection of sentence bert \cite{reimers-2019-sentence-bert} embeddings. 

\subsection{\tool APIs}
\label{sec:backend}

\paragraph{Project management}
The management module provides various interfaces to configure and monitor the annotation project through the \texttt{Project} class. 
\texttt{import\_data} loads the data records from CSV or JSON files into the database. \texttt{set\_config} updates the project configuration as it evolves. \texttt{set\_meta} assigns metadata (e.g.,POS tags, document embeddings) for each data record through user-defined functions. 
\texttt{get\_status} returns the status of the project such as the number of annotated data records and detailed statistic about each label.

A critical feature of \tool is to select interesting subsets of data to show in the widget. 
Subsets can be generated in a user-initiative way via our search engine or a data-driven way via automated suggestions.

\paragraph{Search for subsets}
\tool supports sophisticated searches over data records, annotation, and user-defined metadata through the \texttt{Project.search} API. Users can search data records by keywords (e.g., documents mentioning ``customer service'') or regular expressions to express more complex patterns. The users can also search the database based on already assigned labels (e.g., records with a positive sentiment label). As will be explained later, \tool acknowledges the value of auxiliary information for ML projects and provides advanced search functionalities over metadata. For example, users can query with patterns combining regex expressions and POS tags like \texttt{project.search("(best|amazing) <ADJ> <NOUN>", by="POS")}. 

\paragraph{Automated subset suggestion}
Searches initiated by users can help users explore the dataset in a controlled way, but the quality of searches is only as good as users' knowledge or heuristic about the data and domain. \tool provides an automated subset suggestion engine to assist the exploration. Users can customize the engine by plugging in external suggestion models as needed. Currently, the engine provides two types of techniques:

\begin{itemize}
	\item \textbf{Embedding-based suggestions} makes suggestions based on data embedding vectors provided by the user. \texttt{Subset.suggest\_similar} suggests neighbors of data in the querying subset. \texttt{Project.suggest\_coverage} examines all the data records within the embedding space in an unsupervised way and suggests data points from the less annotated regions to improve annotation coverage of the corpus. 

	\item \textbf{Active suggestions} utilizes active learning techniques to recommend the most informative data for the downstream model. With libraries like ModAL~\cite{modAL2018}, users can select from various selection strategies based on model uncertainty and entropy, etc. Since \tool's seamless notebook experience covers the whole loop from annotation to model training and debugging, users can actively select a subset, annotate the subset, update the model, and test again in the same notebook without switching environments.
\end{itemize}

The output of selection engines are instances of the \texttt{Subset} class with the following methods: \texttt{show} returns a notebook widget for interactive exploration and annotation. \texttt{batch\_annotation} sets the same record-level labels for all data records in the subset. \texttt{suggest\_similar} returns a new subset of the database containing the most similar data for each record in the querying subset according to some metadata with valid distance functions.

\section{Use Case: Sentiment Analysis}
\label{sec:usecase}
We present a use case to illustrate how \tool can support data annotation in NLP researchers and practitioners' workflow.
\username is a data scientist who wants to train a sentiment-related model for her project.
She obtains a Twitter dataset\footnote{The Kaggle dataset~\cite{kaggledata} has ground-truth sentiment labels available. But for demonstration purposes, we ignore them and assume \username only gets the raw Tweets. The dataset contains 14K tweets about major US airlines scraped in February 2015.}about US airlines and decides to label it using \tool. 

\paragraph{Import data} 
In an empty notebook, \username starts by initializing the project named ``Tweet Sentiment'' using a \tool Python API.
She gets a copy of the data from the product manager who often uses Google spreadsheet and imports the data from its published link.

\begin{lstlisting}[language=Python]
from meganno import Project
L = Project(<auth>, "Tweet Sentiment")
L.import_data(<doc_url>, format="csv")
\end{lstlisting}

\paragraph{Set up initial schema}
Without knowing much about the data, \username decides to start the project by collecting binary labels and setting up the project's schema. 
Knowing that \tool supports flexible and editable schema, \username does not worry about getting the perfect schema in the first round and can start exploring and annotating.

\begin{lstlisting}[language=Python]
label_schema = [{
    "label_name": "sentiment_label",
    "level": "record",
    "options": ["Positive", "Negative"]
}]
L.set_config(config1)
\end{lstlisting}

\paragraph{Explore and annotate}

\begin{figure}[t]
\includegraphics[width=\columnwidth]{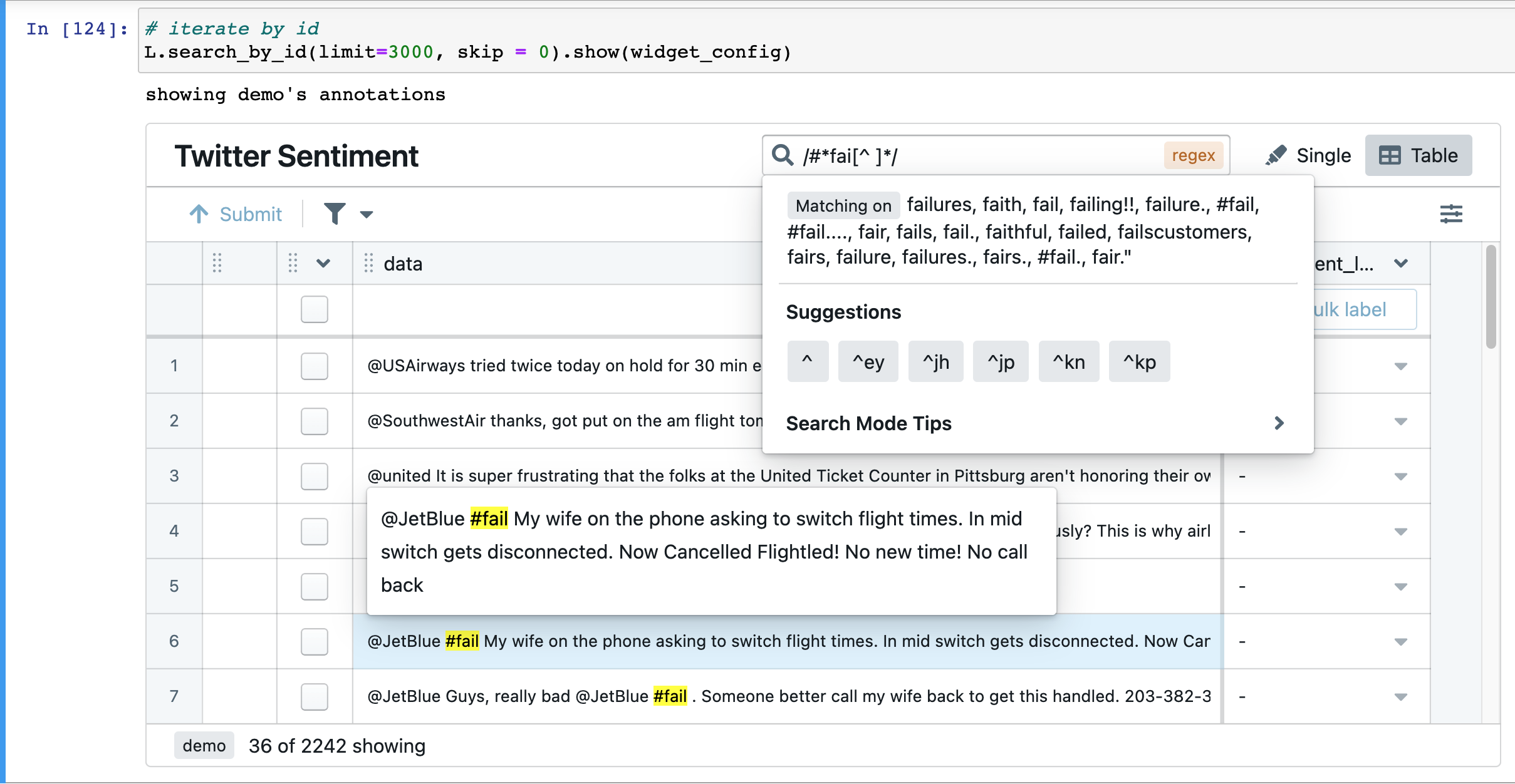}
\caption{UI Search by regular expression. Matched keywords are highlighted.}
\label{fig:widget_usecase_fail}
\end{figure}

She starts by exploring the first 300 data points in the widget's table view.
Using the search box, she filters the subset with the keyword ``amazing''. 
As expected, most of the data records reflect a positive sentiment, so she 
assigns a positive label to multiple data items using the ``Bulk label'' button. 
Next, she wants to examine tweets with hashtags related to failing, so she tries regular expression search using \verb|#fai[^ ]*|~(Fig.~\ref{fig:widget_usecase_fail}).
With better understanding of the dataset, \username chooses to perform more advanced exploration using part-of-speech metadata.
She imports a POS tagger from spaCy~\cite{spacy2} and retrieves tweets that match interesting patterns such as \verb|best <NOUN> of <NOUN>|.

\begin{lstlisting}[language=Python]
import spacy
tagger = spacy.load("en_core_web_sm")

pos1 = L.search("(best|amazing) <ADJ> <NOUN>", by="POS", tagger=tagger)
pos2 = L.search("best <NOUN> of <NOUN>", by="POS", tagger=tagger)
meganno.union(pos1, pos2).show()
\end{lstlisting}
With such an exploratory approach and the batch labeling feature, she can annotate much faster and in a more controlled way.

\paragraph{Candidate suggestion}
After a few more heuristics, she runs out of ideas, so she takes advantage of the suggestion feature. 
She selects the sentence bert~\cite{reimers-2019-sentence-bert} encoder as the meta-data generation function.
She first issues a query with strategy \texttt{similarity} to collect data points similar to the ones retrieved from the previous POS search. 
Finally, she wants to see samples from less covered areas in the embedding space to improve the diversity of the training data and issues a \texttt{coverage} query.

\begin{lstlisting}[language=Python]
m = SentenceTransformer("all-MiniLM-L6")
# set metadata generation function 
L.set_meta("bert", lambda x: list(m.encode(x)))
# get more data like the query result in the previous query(subset_pos)
subset_sim = pos2.suggest_similar(meta_name="bert")
# get data from less covered areas in the embedding space.
subset_cov = L.suggest_coverage(meta_name="bert")
\end{lstlisting}

\paragraph{Update the schema}
\username is now happy with the collected labels, but she aims to go one step further to understand which words or phrases lead to the sentiment judgment. 
So she updates her schema by adding a new span-level label called \texttt{sentiment\_span} with label options being ``happy'' or ``sad''. 
Each data record can have arbitrary numbers of such span-level labels.

\begin{lstlisting}[language=Python]
label_schema = [{
    "label_name": "sentiment_span",
    "level": "span",
    "options": ["Happy", "Sad"] 
}, ... {# previous label options
}]
\end{lstlisting}

After updating the schema, she fetches all records with neutral labels in a widget. 
To highlight and annotate spans, she goes into the single view (as shown in Fig.~\ref{fig:widget_single_sys}). At any step of the iteration, she could refer to the dashboard widget to monitor the project progress. After several rounds of similar iterations, she feels good and concludes her exploration. Finally, \username can export the annotated data in JSON or CSV formats for training or plug in the model directly.

In conclusion, with \tool, \username can explore her dataset using various heuristic-based or automated search functions and better understand the data corpus as she labels. She has the flexibility to iteratively update her schema as the project evolves. Using the widget, \username can finish the entire ML life cyle in the same Jupyter notebook.

\section{Related Work}
\label{sec:relwork}
There exist numerous annotation tools that can support NLP tasks, which are extensively surveyed by ~\citet{neves2019survey,neves2020annotationsaurus}.
In this section, we focus on works that are closer to our flexible, exploratory, efficient, and seamless framework.

\paragraph{Flexible schema}
Unfortunately, most of existing tools are not designed for iterative schema development, and thus they are not flexible enough for evolving projects.
\citet{Felix18} and \citet{Kulesza14} allow users to progressively define document classes by inspecting documents that are assigned to each class.
But these works are more similar to interactive topic modeling or interactive classification, where users assign documents to classes, than document labeling, where users assign labels to documents.
Our tool goes beyond document label refinement and supports a broader task of progressively defining annotation schema (e.g., additionally collecting span-level labels).

\paragraph{Exploratory labeling}
The concept of exploratory labeling is introduced by \citet{Felix18} as using computational techniques to help users group documents into evolving labels.
In our paper, we use the term ``exploratory labeling'' to refer to where exploratory data analysis and data annotation are iteratively conducted in the data understanding/exploration loop.
Exploratory labeling can be beneficial because while labeling data, users gain insight into their dataset~\cite{Sun17}.

\paragraph{Efficient batch/bulk annotation}
A few tools offer a functionality to simultaneously assign labels to multiple spans within a record.
For example, YEDDA~\cite{yang2018yedda} can annotate multiple span-level labels via command line.
TALEN~\cite{talen2018}, a named entity tagging tool, has an entity propagation feature which annotates all mentions of an entity in a document at once.
In contrast, users can annotate multiple records simultaneously using our Python API and a GUI widget.

\paragraph{Notebook widget}
Computational notebooks
are frequently used by data analysts to iteratively write and edit code to understand data, test hypotheses, and build models~\cite{head2019,randles2017notebook}.
Following the practice of mage~\cite{mage-fluid-moves-between-code} which extends Jupyter notebook with GUI widgets for specific tasks, our widget is designed to achieve flexible communication with the rest of ML development codes.
Annotation tools which are implemented as Jupyter widgets include Pigeon~\cite{pigeon} and ipyannotate~\cite{ipyannotate}, but they only offer a simple label assignment feature.

\section{Conclusion}
\label{sec:conclusion}
In this paper, we present \tool, an annotation framework designed for NLP researchers and practitioners.
Through \tool's programmatic interfaces and interactive widget, users can iteratively explore and search for interesting data subsets, annotate data, train models, evaluate and debug models within a Jupyter notebook without the overhead of context switch or data migration. 

\bibliography{ref}

\begin{thebibliography}{23}
\expandafter\ifx\csname natexlab\endcsname\relax\def\natexlab#1{#1}\fi

\bibitem[{hua()}]{huamnloop}

\newblock humanloop.com.
\newblock \url{https://humanloop.com/}.

\bibitem[{ipy()}]{ipyannotate}

\newblock ipyannotate.
\newblock \url{https://github.com/ipyannotate/ipyannotate}.

\bibitem[{pig()}]{pigeon}

\newblock pigeon.
\newblock \url{https://github.com/agermanidis/pigeon}.

\bibitem[{Danka and Horvath()}]{modAL2018}
Tivadar Danka and Peter Horvath.
\newblock \href {https://github.com/cosmic-cortex/modAL} {mod{AL}: {A} modular
  active learning framework for {P}ython}.
\newblock Available on arXiv at \url{https://arxiv.org/abs/1805.00979}.

\bibitem[{Felix et~al.(2018)Felix, Dasgupta, and Bertini}]{Felix18}
Cristian Felix, Aritra Dasgupta, and Enrico Bertini. 2018.
\newblock \href {https://doi.org/10.1145/3242587.3242596} {The exploratory
  labeling assistant: Mixed-initiative label curation with large document
  collections}.
\newblock In \emph{Proceedings of the 31st Annual ACM Symposium on User
  Interface Software and Technology}, UIST '18, pages 153--164, New York, NY,
  USA. Association for Computing Machinery.

\bibitem[{for Everyone~library()}]{kaggledata}
Crowdflower~Data for Everyone~library.
\newblock \href {https://www.kaggle.com/crowdflower/twitter-airline-sentiment}
  {{Twitter US Airline Sentiment(Version 4)}}.

\bibitem[{Geiger et~al.(2021)Geiger, Cope, Ip, Lotosh, Shah, Weng, and
  Tang}]{garbage2021}
R.~Stuart Geiger, Dominique Cope, Jamie Ip, Marsha Lotosh, Aayush Shah, Jenny
  Weng, and Rebekah Tang. 2021.
\newblock \href {https://doi.org/10.1162/qss_a_00144} {"garbage in, garbage
  out" revisited: What do machine learning application papers report about
  human-labeled training data?}
\newblock \emph{Quantitative Science Studies}, pages 1--33.

\bibitem[{Head et~al.(2019)Head, Hohman, Barik, Drucker, and DeLine}]{head2019}
Andrew Head, Fred Hohman, Titus Barik, Steven~M. Drucker, and Robert DeLine.
  2019.
\newblock \href {https://doi.org/10.1145/3290605.3300500} {\emph{Managing
  Messes in Computational Notebooks}}, pages 1--12. Association for Computing
  Machinery, New York, NY, USA.

\bibitem[{Hohman et~al.(2020)Hohman, Wongsuphasawat, Kery, and
  Patel}]{hohman2020iterativeML}
Fred Hohman, Kanit Wongsuphasawat, Mary~Beth Kery, and Kayur Patel. 2020.
\newblock Understanding and visualizing data iteration in machine learning.
\newblock In \emph{Proceedings of the 2020 CHI Conference on Human Factors in
  Computing Systems}, pages 1--13.

\bibitem[{Honnibal and Montani(2017)}]{spacy2}
Matthew Honnibal and Ines Montani. 2017.
\newblock {spaCy 2}: Natural language understanding with {B}loom embeddings,
  convolutional neural networks and incremental parsing.
\newblock To appear.

\bibitem[{Kery et~al.(2020)Kery, Ren, Hohman, Moritz, Wongsuphasawat, and
  Patel}]{mage-fluid-moves-between-code}
Mary~Beth Kery, Donghao Ren, Fred Hohman, Dominik Moritz, Kanit Wongsuphasawat,
  and Kayur Patel. 2020.
\newblock \href {https://arxiv.org/pdf/2009.10643.pdf} {mage: Fluid moves
  between code and graphical work in computational notebooks}.

\bibitem[{Kulesza et~al.(2014)Kulesza, Amershi, Caruana, Fisher, and
  Charles}]{Kulesza14}
Todd Kulesza, Saleema Amershi, Rich Caruana, Danyel Fisher, and Denis Charles.
  2014.
\newblock \href {https://doi.org/10.1145/2556288.2557238} {Structured labeling
  for facilitating concept evolution in machine learning}.
\newblock In \emph{Proceedings of the SIGCHI Conference on Human Factors in
  Computing Systems}, CHI '14, pages 3075--3084, New York, NY, USA. Association
  for Computing Machinery.

\bibitem[{Montani and Honnibal(2018)}]{Prodigy:2018}
Ines Montani and Matthew Honnibal. 2018.
\newblock \href {http://arxiv.org/abs/to appear} {Prodigy: A new annotation
  tool for radically efficient machine teaching}.
\newblock \emph{Artificial Intelligence}, to appear.

\bibitem[{Neves and Seva(2019)}]{neves2019survey}
Mariana Neves and Jurica Seva. 2019.
\newblock \href {https://doi.org/10.1093/bib/bbz130} {{An extensive review of
  tools for manual annotation of documents}}.
\newblock \emph{Briefings in Bioinformatics}, 22(1):146--163.

\bibitem[{Neves and Seva(2020)}]{neves2020annotationsaurus}
Mariana Neves and Jurica Seva. 2020.
\newblock \href {http://arxiv.org/abs/2010.06251} {Annotationsaurus: A
  searchable directory of annotation tools}.

\bibitem[{Pustejovsky and Stubbs(2012)}]{pustejovsky2012natural}
James Pustejovsky and Amber Stubbs. 2012.
\newblock \emph{Natural Language Annotation for Machine Learning: A guide to
  corpus-building for applications}.
\newblock " O'Reilly Media, Inc.".

\bibitem[{Rahman and Kandogan(2022)}]{rahman2022ie}
Sajjadur Rahman and Eser Kandogan. 2022.
\newblock \href {https://doi.org/10.1145/3491102.3502068} {Characterizing
  practices, limitations, and opportunities related to text information
  extraction workflows: A human-in-the-loop perspective}.
\newblock CHI '22, New York, NY, USA. Association for Computing Machinery.

\bibitem[{Randles et~al.(2017)Randles, Pasquetto, Golshan, and
  Borgman}]{randles2017notebook}
Bernadette~M. Randles, Irene~V. Pasquetto, Milena~S. Golshan, and Christine~L.
  Borgman. 2017.
\newblock \href {https://doi.org/10.1109/JCDL.2017.7991618} {Using the jupyter
  notebook as a tool for open science: An empirical study}.
\newblock In \emph{2017 ACM/IEEE Joint Conference on Digital Libraries (JCDL)},
  pages 1--2.

\bibitem[{Reimers and Gurevych(2019)}]{reimers-2019-sentence-bert}
Nils Reimers and Iryna Gurevych. 2019.
\newblock \href {https://arxiv.org/abs/1908.10084} {Sentence-bert: Sentence
  embeddings using siamese bert-networks}.
\newblock In \emph{Proceedings of the 2019 Conference on Empirical Methods in
  Natural Language Processing}. Association for Computational Linguistics.

\bibitem[{Stephen~Mayhew(2018)}]{talen2018}
Dan~Roth Stephen~Mayhew. 2018.
\newblock Talen: Tool for annotation of low-resource entities.
\newblock In \emph{ACL System Demonstrations}.

\bibitem[{Sun et~al.(2017)Sun, Lank, and Terry}]{Sun17}
Yunjia Sun, Edward Lank, and Michael Terry. 2017.
\newblock \href {https://doi.org/10.1145/3025171.3025208} {Label-and-learn:
  Visualizing the likelihood of machine learning classifier's success during
  data labeling}.
\newblock In \emph{Proceedings of the 22nd International Conference on
  Intelligent User Interfaces}, IUI '17, pages 523--534, New York, NY, USA.
  Association for Computing Machinery.

\bibitem[{Van~der Maaten and Hinton(2008)}]{van2008visualizing}
Laurens Van~der Maaten and Geoffrey Hinton. 2008.
\newblock Visualizing data using t-sne.
\newblock \emph{Journal of machine learning research}, 9(11).

\bibitem[{Yang et~al.(2018)Yang, Zhang, Li, and Li}]{yang2018yedda}
Jie Yang, Yue Zhang, Linwei Li, and Xingxuan Li. 2018.
\newblock \href {https://doi.org/10.18653/v1/P18-4006} {{YEDDA}: A lightweight
  collaborative text span annotation tool}.
\newblock In \emph{Proceedings of {ACL} 2018, System Demonstrations}, pages
  31--36, Melbourne, Australia. Association for Computational Linguistics.

\end{thebibliography}
\bibliographystyle{acl_natbib}

\end{document}